\documentclass[aps,prd,12pt,twocolumn,superscriptaddress,showpacs,showkeys,nofootinbib,reprint,longbibliography,amsmath,amsfonts,amssymb]{revtex4-1}
\usepackage{siunitx}
\usepackage{tikz-cd}
\usepackage{ytableau}
\ytableausetup{aligntableaux=center,boxsize=1em}
\usepackage[colorlinks,allcolors=blue]{hyperref}

\begin{document}

\title{A model for cosmological perturbations in affine gravity}

\author{Oscar Castillo-Felisola}
\email[Corresponding author: ]{o.castillo.felisola:at:proton.me}
\affiliation{Departamento de F\'isica, Universidad T\'ecnica Federico Santa Mar\'\i a\\ Casilla 110-V, Valpara\'iso, Chile}

\author{Radouane Gannouji}
\email{radouane.gannouji@pucv.cl}
\affiliation{Instituto de Física, Pontificia Universidad Católica de Valparaíso\\ Av. Brasil 2950, Valparaíso, Chile}

\author{Manuel Morocho-L\'opez}
\email{manuel.morocho@usm.cl}
\affiliation{Departamento de F\'isica, Universidad T\'ecnica Federico Santa Mar\'\i a\\ Casilla 110-V, Valpara\'iso, Chile}
\affiliation{Instituto de Física, Pontificia Universidad Católica de Valparaíso\\ Av. Brasil 2950, Valparaíso, Chile}
\affiliation{Facultad de Ciencias Naturales y Matem\'aticas, Escuela Superior Polit\'ecnica del Litoral\\ Km. 30.5 V\'ia Perimetral, Guayaquil, Ecuador}

\author{Marcelo Rozas-Rojas}
\email{marcelo.rozas@pucv.cl}
\affiliation{Departamento de F\'isica, Universidad T\'ecnica Federico Santa Mar\'\i a\\ Casilla 110-V, Valpara\'iso, Chile}
\affiliation{Instituto de Física, Pontificia Universidad Católica de Valparaíso\\ Av. Brasil 2950, Valparaíso, Chile}

\date{\today}

\begin{abstract}
In this paper, we present the cosmological perturbation formalism for theories within the framework of affine gravity. These theories are distinguished by their connection, devoid of any metric. Our approach involves segregating perturbations into symmetric and antisymmetric components (related to torsion), each further decomposed into irreducible elements, namely scalars, pseudoscalars, vectors, pseudovectors, 2-tensors, and 3-tensors. Finally, we have fully addressed the gauge freedom in this context.  
\end{abstract}

\maketitle

\section{Introduction}
\label{sec:introduction}
The cosmological model of the Universe is built up under the precept of isotropy and homogeneity, and assumes that the theory that explains the gravitational interactions is General Relativity. Such premise is often called the \emph{cosmological principle}, and it is understood as valid in terms of averages over \emph{large} portions of the space. It was soon understood that keeping these condition would not explain the structure formation on the observable Universe.

The structure formation then requires to departure from the cosmological principle. Given the complexity of the Einstein equations, even in the simpler scenarios, the strategy is to consider perturbations around an isotropic and homogeneous background, and find the conditions imposed by the field equations on the perturbations (up to a certain order). In General Relativity the metric is the field responsible of mediating the gravitational interaction, hence one considers that the physical metric, \(g_{\mu\nu}\), is a sum of an isotropic and homogeneous background metric \(\bar{g}_{\mu\nu}\) plus a (small) symmetric tensor field \(h_{\mu\nu}\),
\begin{equation}
  g_{\mu\nu} = \bar{g}_{\mu\nu} + \epsilon h_{\mu\nu},
  \label{eq:metric-perturbation}
\end{equation}
where \(\epsilon\) is an expansion parameter that allows to control the order of the perturbation.

Moreover, within the framework of the \(\Lambda\)CDM model (\(\Lambda\) cold dark matter, where the \(\Lambda\) stands for the existence of a cosmological constant) the observed Universe is composed mainly by \emph{dark energy} (69\%) and \emph{dark matter} (26\%) \cite{Planck:2018vyg}, which together constitute the so called dark sector of the Universe, two forms of matter whose origin and nature is still not well understood, nor have been observed directly. Indeed, we have been able to observe the effects produced by their possible existence, but without encountering a direct detection such as in a particle accelerator.

The necessity of a huge dark sector to comprehend the Universe within the foundations of the \(\Lambda\)CDM model has posed the question of whether our theories of gravity and particles are fundamental or just effective manifestations of yet-to-be-discovered fundamental ones. This has driven the formulation of uncountable generalisations of our standard models, with the addition of \emph{exotic}  types of matter and interactions from the particle physics point of view, and the extension of the gravitational sector from the gravitational counterpart.

Focusing on the branch of extensions of the gravitational sector, one encounters for example metric models of gravity which consider higher order in curvature (e.g. Lanczos--Lovelock gravities \cite{Lanczos:1938sf,lovelock69,Lovelock:1971yv,lovelock89_tensor}), but there are also metric-affine models of gravity (see Refs. \cite{Hehl:1994ue,Sotiriou:2006qn,Capozziello:2009mq} for detailed reviews), in which the metric and the affine connection are considered as independent fields. A modern review of many generalised gravitational sector can be found in \cite{CANTATA:2021ktz} (see also \cite{Gannouji:2019mph} for a student friendly introduction). In addition, there exist purely affine models of gravity \cite{einstein23,einstein23_theor_affin_field,eddington23,schroedinger50_space,kijowski78,Kijowski:2004bj,Krasnov:2007ei,Krasnov:2007ky,Krasnov:2007uu,Krasnov:2011pp,Poplawski:2007ik,Poplawski:2012bw,Skirzewski:2014eta,Castillo-Felisola:2015cqa}, in which the mediator of the gravitational interaction comes from the affine connection, and the metric tensor field is not required to build up the action functional of the model.

In this later construction of gravity, cosmological solutions have been studied in \cite{Azri:2017uor,Shimada:2018lnm,Iosifidis:2019dua,Castillo-Felisola:2019wcs,Iosifidis:2021fnq,Capozziello:2022vyd,Iosifidis:2023but}. Nonetheless, the analysis of perturbations around an exact solution requires to consider perturbations of the affine connection. However, to the extent of our knowledge the perturbative analysis of an affine connection has not been reported in the literature.

The aim of this paper is to study\footnote{while completing this work, a paper on similar topic but with a different approach has been published \cite{Aoki:2023sum}} the cosmological perturbations for the affine connection, which might serve as tool to study the inhomogeneities of the Universe in affine and metric-affine models of gravity.

In the first part of this paper, we summarize some notions of differential geometry focusing on the objects that we will use in this paper. In the second part, we will give the different components of the connection compatible with the cosmological principle followed by the complete study of perturbations. Finally, we will address the problem of gauge freedom and construct gauge invariant quantities as well as list the different possible gauges. Finally, in the last section we conclude the paper and state some possible future directions of this work.

\section{Brief review of differential geometry}
\label{sec:differential-geometry}
The content of this section is based on Refs. \cite{eisenhart27_non_rieman,nomizu56_lie,choquet-bruhat89_analy,joyce00_compac_manif_special_holon,nakahara05_geomet_topol_physic,besse07_einst,schouten13_ricci}.

An \(m\)-dimensional differential manifold, \(M\), is a topological space which is locally homeomorphic to \(\mathbb{R}^n\) with a differential structure defined by the transition functions.\footnote{For every open ball \(U_i \in M\) there exists a continuous map \(\phi_i\) with continuous inverse \(\phi^{-1}_i\) such that on the overlap of two open balls, \(U_i \cap U_j \neq \emptyset\), the transition functions \(\psi_{ji} \equiv \phi_j \circ \phi^{-1}_i \colon \mathbb{R}^m \to \mathbb{R}^m\) are \(C^k(\mathbb{R}^m)\). Customarily, if the order of differentibility (\(k\)) is not specified, it is assumed that the transition functions are \(C^{\infty}(\mathbb{R}^m)\).}

The differential structure allows to define at each point of the manifold (\(m \in M\)) a vector tangent space, \(T_m M\). Using the tensor product one defines the tensor space \(\otimes^p T_m M\) over the point \(m\), where the \(p\)-times contra-variant tensors live. The set of linear maps from \(T_m M\) to the algebraic field \(\mathbb{R}\) form a vector space dubbed vector cotangent space, \(T^{ * }_m M\), and the tensor product \(\otimes^q T^{ * }_m M\) defines the space where the \(q\)-times covariant tensor live. Clearly, it is possible to define the space \(\otimes^p T_m M \otimes^q T^{ * }_m M\) where the \((p,q)\)-tensors live.

The disjoint union of the vector tangent spaces over \(M\) defines the tangent bundle \(TM\), which is a \(2m\)-dimensional differential manifold, whose sections are the vector fields on \(M\). A similar construction can be made with the cotangent spaces yields the cotangent bundle, \(T^{ * }M\), whose sections are the covector (one-form) fields on \(M\). The \((p,q)\)-tensor fields are sections on the bundle \(\otimes^p TM \otimes^q T^{ * }M\) over \(M\). A sub-bundle \(\Lambda^q T^{ * }M \subset \otimes^q T^{ * }M\) over \(M\) is the bundle whose sections are \(q\)-forms, i.e. completely skew-symmetric \((0,q)\)-tensor fields on \(M\).\footnote{The bundle \(\Lambda^{q} T^{ * }M\) can be understood as the \(q^{\mathrm{th}}\) exterior power of the cotangent bundle.} A volume form, \(\omega\), is a nowhere vanishing section of the bundle \(\Lambda^m T^{ * }M\), and this is the structure that allows us to define the integration on the manifold \(M\).

In order to define derivatives of the geometrical objects on \(M\), e.g. tensor fields, a structure that allows to compare the object placed at different points of \(M\) is required. Such structure is called a linear connection (in this article we make no distinction between linear and affine connections, so hereon we shall refer indistinctly to it as the affine connection), \(\nabla\), and it is determined by its components, \(\Gamma_{\mu}{}^{\lambda}{}_{\rho}\). The connection contains the information of how the vector basis changes as one \emph{moves} on the manifold \(M\), e.g. if we consider a vector basis \(\{\vec{e}_{\rho}\}\) the components of the connection are defined as
\begin{equation}
  \partial_{\mu} \vec{e}_{\rho} = \Gamma_{\mu}{}^{\lambda}{}_{\rho} \vec{e}_{\lambda}.
  \label{eq:def-connection}
\end{equation}
The definition from Eq. \eqref{eq:def-connection} induces an action of the connection over the tensor bundles of any type, and hence the sections whose action of the affine connection is well defined (at least over a region) are said to be differential section. In particular, infinitely differentiable sections are called smooth sections, and the set of smooth sections is denoted by \(C^{\infty}(\otimes^p TM \otimes^q T^{ * }M)\).

The affine connection serves to define the notion of parallelism, and as a tool it might be interpreted as the analogous to the \emph{ruler} in classical geometry. In addition, the notion of curvature tensor stands solely on the existence of this structure. Hence, the notion of curvature,
\begin{equation}
  \label{eq:def-curvature-formal}
  R(X,Y) Z = \nabla_X \nabla_Y Z - \nabla_Y \nabla_X Z - \nabla_{[X,Y]} Z,
\end{equation}
with \({X,Y,Z} \in C^{\infty}(TM)\), which can be written in components as
\begin{equation}
  \label{eq:def-curvature}
  {\mathcal{R}}_{\mu\nu}{}^{\lambda}{}_{\rho}
  =
  \partial_{\mu} \Gamma_{\nu}{}^{\lambda}{}_{\rho} -
  \partial_{\nu} \Gamma_{\mu}{}^{\lambda}{}_{\rho} +
  \Gamma_{\mu}{}^{\lambda}{}_{\sigma} \Gamma_{\nu}{}^{\sigma}{}_{\rho} -
  \Gamma_{\nu}{}^{\lambda}{}_{\sigma} \Gamma_{\mu}{}^{\sigma}{}_{\rho},
\end{equation}
makes sense in \emph{affinely connected manifolds}, which are manifolds endowed with an affine structure, \((M,\nabla)\). These affinely connected manifolds are the arena were purely affine gravitational model are built up. 

With a generic affine connection one could define another tensor quantity called the torsion,
\begin{equation}
  T(X,Y) = \nabla_X Y - \nabla_Y X - [X,Y],
  \label{eq:def-torsion-formal}
\end{equation}
for \({X,Y} \in C^{\infty}(TM)\), which in components is expressed like
\begin{equation}
  {\mathcal{T}}_{\mu}{}^{\lambda}{}_{\nu} = \Gamma_{\mu}{}^{\lambda}{}_{\nu} - \Gamma_{\nu}{}^{\lambda}{}_{\mu}.
  \label{eq:def-torsion}
\end{equation}

A metric tensor field is a differentiable section of the symmetric sub-bundle \(S^2(T^{ * }M) \subset \otimes^2 T^{ * }M\), \(g \in C^{\infty}(S^2(T^{ * }M))\), which is non-degenerated at each point \(p \in M\), i.e. at any \(p\) if \(g_p(X,Y) = 0\) for every \(Y \in T_p M\) then necessarily \(X = 0\). Every metric tensor fields has a property called signature, which correspond to the sign of its \emph{eigenvalues}, that has to be consistent all over the manifold. The metric tensor is the tool that define (generalised) distances, and hence an analogous to the compass in classical geometry.

The metric provides a map from \(C^{\infty}(TM)\) to \(C^{\infty}(T^{ * }M)\), and its inverse (also called dual) is a map in the opposite direction. This two objects define the \emph{musical} maps, which are said to lower and raise the indices.

A metric-affine manifold is a differentiable manifold equipped with an affine connection and a metric, \((M,\nabla,g)\). Note that in general there is no relation between the affine and metric structures. However, once the two structures are given, the affine connection can be decomposed into irreducible parts compatible with the metric field as follow: (i) calculate the covariant derivative of the metric to define the non-metricity tensor field \(\mathcal{Q}_{\mu\nu\lambda} = \nabla_{\mu} g_{\nu\lambda}\); (ii) use the metric to lower the index of the torsion, and obtain the triple covariant torsion \(\mathcal{T}_{\mu\nu\lambda} = g_{\nu\sigma} {\mathcal{T}}_{\mu}{}^{\sigma}{}_{\lambda}\); (iii) define the operation
\begin{equation}
  \psi_{\{\mu\nu\lambda\}} = \psi_{\mu\nu\lambda} - \psi_{\nu\lambda\mu} + \psi_{\lambda\mu\nu}.
  \label{eq:def-alternating-operation}
\end{equation}
Then, connection can be written as
\begin{equation}
  \Gamma_{\mu}{}^{\lambda}{}_{\rho} = \frac{1}{2} g^{\lambda\sigma} \left( \partial_{\{\mu} g_{\sigma\rho\}} + \mathcal{T}_{\{\mu\sigma\rho\}} + \mathcal{Q}_{\{\mu\sigma\rho\}} \right),
  \label{eq:parts-of-connection}
\end{equation}
where the first term on the right-hand side corresponds to the Levi-Civita connection (which is defined solely by the metric tensor field, and is the connection that appears in metric models of gravity like General Relativity), the second term is known as the contorsion (contains the information about the torsion), and the third one is called disformation (encodes the information of the non-metricity). The joint contribution of the contorsion and disformation is often refer to as the distorsion.

According with the decomposition of the affine connection, the affinely connected manifolds can be classified into classes as depicted in the following commutative diagram,
\begin{equation}
  \begin{tikzcd}[row sep=scriptsize, column sep=scriptsize]
    & (M,{\Gamma}) \ar[d, "g"] & & \\
    & (\mathcal{Q},\mathcal{R},\mathcal{T}) \arrow[dl] \arrow[rr] \arrow[dd] & & (\mathcal{Q},\mathcal{T}) \arrow[dl] \arrow[dd] \\
    (\mathcal{Q},\mathcal{R}) \arrow[rr, crossing over] \arrow[dd] & & (\mathcal{Q}) \\
    & (\mathcal{R},\mathcal{T}) \arrow[dl] \arrow[rr] & & (\mathcal{T}) \arrow[dl] \\
    (\mathcal{R}) \arrow[rr] & & (\text{Flat}) \arrow[from=uu, crossing over]\\
  \end{tikzcd}
  \label{eq:type-of-manifolds}
\end{equation}

Therefore, from the commutative diagram in Eq. \eqref{eq:type-of-manifolds} one reads that affinely connected manifolds \((M,\Gamma)\), when endowed with a metric are characterised by three quantities, the curvature (\(\mathcal{R}\)), the torsion (\(\mathcal{T}\)) and the non-metricity (\(\mathcal{Q}\)). Particular cases are those in whose some of these quantities vanish, for example manifolds that posses solely curvature are said to be Riemannian, others characterised by their torsion are called Weitzenböck, those that posses curvature and torsion are known as Riemann--Cartan, etc \cite{Hehl:1994ue,Castillo-Felisola:2023pdx}.

Before ending this section it is worth to highlight that the logical order of structures in differential geometry is not the one presented in most textbooks of General Relativity.\footnote{Schrödiger presents it in the right order in his book \cite{schroedinger50_space}.} Therefore, the most fundamental structure is the exterior algebra (induced by the tensor bundle), which defines a volume, and hence allows to define integration on manifolds. With this first structure one could in principle define action functionals. Then, a second structure is the affine structure, which allows to compare geometrical objects based at different point of the manifold. This structure yields the covariant differentiation, and allows us to define parallelism, and particularly parallel transport and auto-parallel curves. Finally, a third structure is the metric structure, which allows to define a consistent way to measure all over the manifold.\footnote{It is possible to give a notion of \emph{distance} along a given auto-parallel curve using the affine structure. However, that notion cannot be extended consistently to other auto-parallel curves.}

\section{Cosmological principle and geometrical objects}
\label{sec:cosmological-principle}
In this section we restrict ourselves to four-dimensional manifolds, and we assume that the coordinate system is \emph{spherical} with coordinates \(x^{\mu} = (t, r, \theta, \varphi)\). Most of the development of the section would not require to specify details of the signature of the metric, but it is safe to assume that our signature is positive, i.e. most of the signs (if not all) are positive.

The cosmological principle requires isotropy and homogeneity along a three-dimensional submanifold, which in our case would be parametrised by the coordinates \(x^i = (r, \theta, \varphi)\).

Isotropy ensures that our three-dimensional submanifold is indistinguishable under rotations, while homogeneity ensures that under translation of the \emph{origin} of coordinates the submanifold is also indistinguishable. Clearly, the cosmological principle associates a \(6\)-dimensional (continuous) symmetry group to the three-dimensional submanifold. There are three possible choice of the symmetry group: (i) the Euclidean group \(E_3\); (ii) the orthogonal group \(SO(4)\); or (iii) the orthogonal group \(SO(3,1)\).\footnote{The group \(SO(2,2)\) has been left out the list because it would change the signature when acting in the three-dimensional submanifold.} The Killing vectors associated with these groups can be written explicitly as follows,
\begin{align}
  J_1
  & =
    \begin{pmatrix}
      0 & 0 & - \cos \theta & \cot \theta \sin \varphi
    \end{pmatrix},
  \\
  J_2
  & =
    \begin{pmatrix}
      0 & 0 & \sin \theta & \cot \theta \sin \varphi
    \end{pmatrix},
  \\
  J_3
  & =
    \begin{pmatrix}
      0 & 0 & 0 & 1
    \end{pmatrix},
  \\
  P_1
  & = \sqrt{1 - \kappa r^2}
    \begin{pmatrix}
      0 & \sin \theta \cos \varphi & \frac{\cos \theta \cos \varphi}{r} & - \frac{\sin \varphi}{\sin \theta}
    \end{pmatrix},
  \\
  P_2
  & = \sqrt{1 - \kappa r^2}
    \begin{pmatrix}
      0 & \sin \theta \sin \varphi & \frac{\cos \theta \sin \varphi}{r} & - \frac{\cos \varphi}{\sin \theta}
    \end{pmatrix},
  \\
  P_3
  & = \sqrt{1 - \kappa r^2}
    \begin{pmatrix}
      0 & \cos \theta & - \frac{\sin \theta}{r} & 0
    \end{pmatrix},
\end{align}
where the parameter \(\kappa\) is given by
\begin{equation}
  \kappa =
  \begin{cases}
    1 & SO(4) \\
    0 & E_3 \\
    -1 & SO(3,1)
  \end{cases}.
  \label{eq:value-kappa}
\end{equation}

It is well known that an \(n\)-dimensional maximally symmetric space possesses \(n (n + 1) / 2\) Killing vectors \cite{thirring79_cours_mathem_physic_ii}. Therefore, the isotropic and homogeneous three-dimensional submanifold described above is maximally symmetric. Using the Lie derivative it is straightforward to show that on the maximally symmetric submanifold the symmetries determine a metric \(s_{ij}\) and a connection \(\gamma_{i}{}^{j}{}_{k}\) (for a description of the method, see Ref. \cite{Castillo-Felisola:2019thp}). The induced metric is given by
\begin{equation}
  \label{eq:three-metric}
  s_{ij} =
  \begin{pmatrix} 
    \frac{1}{1 - \kappa r^2} & 0 & 0 \\
    0 & r^2 & 0 \\
    0 & 0 & r^2 \sin^2 \theta 
  \end{pmatrix},
\end{equation}
the symmetric part of the connection by
\begin{equation}
  \begin{aligned}
    \gamma_{r}{}^{r}{}_{r} & = \frac{\kappa r}{1 - \kappa r^2},
    &
    \gamma_{\theta}{}^{r}{}_{\theta} & = \kappa r^3 - r,
    \\
    \gamma_{\varphi}{}^{r}{}_{\varphi} & = \left(\kappa r^3 - r\right)\sin
    ^2\theta,
    &
    \gamma_{(r}{}^{\theta}{}_{\theta)} & = \frac{1}{r},
    \\
    \gamma_{\varphi}{}^{\theta}{}_{\varphi} & = - \cos \theta \sin \theta,
    &
    \gamma_{(r}{}^{\varphi}{}_{\varphi)} & =\frac{1}{r},
    \\
    \gamma_{(\theta}{}^{\varphi}{}_{\varphi)} & = \frac{\cos \theta}{\sin
      \theta},
    & &
  \end{aligned}
  \label{eq:three-connection}
\end{equation}
and the additional torsional terms are (up to a multiplicative factor)
\begin{equation}
  {\gamma}_{[i}{}^{j}{}_{k]} \equiv S_{i}{}^{j}{}_{k} = \sqrt{s} \, s^{jl} \varepsilon_{ilk}.
  \label{eq:three-connection-torsional-term}
\end{equation}

The ansätze of the four-dimensional connection is written in terms of the three-dimensional geometrical objects as follows \cite{Castillo-Felisola:2019wcs,Castillo-Felisola:2019thp,Castillo-Felisola:2018jfp},
\begin{equation}
  \label{eq:cosmological-connection}
  \begin{aligned}
    \Gamma_{t}{}^{t}{}_{t} & = f,
    &
    \Gamma_{(i}{}^{t}{}_{j)} & = g \, s_{ij},
    \\
    \Gamma_{(t}{}^{i}{}_{j)} & = h \, \delta^{i}_{j},
    &
    \Gamma_{(i}{}^{j}{}_{k)} & = \gamma_{(i}{}^{j}{}_{k)},
    \\
    \Gamma_{[t}{}^{i}{}_{j]} & = p \, S_{t}{}^i{}_j
    &
    \Gamma_{[i}{}^{j}{}_{k]} & = q \, S_{i}{}^{j}{}_{k},
    \\
    & = - p \, \delta^i_j,
  \end{aligned}
\end{equation} 
where \(f\), \(g\), \(h\), \(p\) and \(q\) are functions of time, which are determined when one requires the field equations to be satisfied (and hence their values are model dependent).

Similarly, it is possible to determine the form of the most general four-dimensional metric compatible with the cosmological principle,
\begin{equation}
  g_{\mu\nu} = N(t)^2 \, \delta^{t}_{\mu} \delta^{t}_{\nu} + a(t)^2 \, s_{ij} \, \delta^{i}_{\mu} \delta^{j}_{\nu}.
  \label{eq:cosmological-metric}
\end{equation}
It is worth to highlight a couple of points here: (i) The metric in Eq. \eqref{eq:cosmological-metric} won't be used in the analysis presented in the following section, and; (ii) The signature of the metric might be encoded in the function \(N\). Interestingly, even if some affine models, like the polynomial affine model of gravity, do not require a pre-existing fundamental metric field for its formulations, it is possible to find connection  \emph{descendent} metrics in the space of solutions.

\section{Analysis of perturbations}
\label{sec:perturbations}
In standard cosmology, which is based on General Relativity (a metric model of gravity), one starts considering an isotropic and homogeneous background, and the departure from the cosmological principle is treated via perturbation theory (see for example Ref. \cite{Mukhanov:1990me,weinberg08_cosmol,poisson14_gravit,dodelson20_moder}).

Mathematically, one proposes that the \emph{expansion} of the metric as in Eq. \eqref{eq:metric-perturbation} is enlarged with a similar expansion for the energy-momentum tensor, from which we obtain the linearized field equations \cite{Wald:1984rg},
\begin{equation}
  \frac{\mathrm{d}}{\mathrm{d}\epsilon} \Big( G_{\mu\nu}(g) \Big) \bigg|_{\epsilon = 0}
  =
  \frac{\mathrm{d}}{\mathrm{d}\epsilon} \Big( T_{\mu\nu}(g,\rho,p) \Big) \bigg|_{\epsilon = 0} .
  \label{eq:def-linearised-feqs}
\end{equation}
Secondly, using the natural splitting between the \emph{time} and spatial coordinates, the metric perturbation is separated in a \((3+1)\)-decomposition as,\footnote{Remind that since the general metric \(g\) has to be symmetric, the tensor field \(h\) is symmetric too. Hence, \(h_{t i} = h_{i t}\).}
\begin{equation*}
  h_{\mu\nu} \to \left\{ h_{t t}, h_{t i}, h_{i j} \right\}.
\end{equation*}
Moreover, each of these parts can decomposed further into longitudinal and transverse components, according to the Helmholtz decomposition, which is written as follows \cite{Lifshitz:1945du,Straumann:1997mtl},\footnote{Beware of the difference between the covariant derivative, \(D_i\), and the tensor component of the decomposition, \(\hat{D}_{ij}\).}
\begin{equation}
  \begin{aligned}
    h_{t t}
    & =
    \hat{E},
    \\
    h_{t i}
    & =
    D_i \hat{F} + \hat{G}_i,
    \\
    h_{i j}
    & =
    \hat{A} \, s_{i j} + D_i D_j \hat{B} + 2 D_{(i} \hat{C}_{j)} + \hat{H}_{i j},
  \end{aligned}
  \label{eq:helmholtz-decomposition-metric}
\end{equation}
where \(D_i\) denotes the three-dimensional Riemannian covariant derivative (i.e. the derivative compatible with the three-dimensional metric field \(s_{ij}\)), the functions \(\hat{A}\), \(\hat{B}\), \(\hat{C}_i\), \(\hat{E}\), \(\hat{F}\), \(\hat{G}_i\) and \(\hat{H}_{ij}\) depend on the whole set of coordinates \(x^{\mu} = (t, \vec{x})\), and also \(\hat{H}_{ij} = \hat{H}_{ji}\), \(\hat{H}^i{}_i = D_i \hat{H}^i{}_j = D_i \hat{C}^i = D_i \hat{G}^i = 0\). In Eq. \eqref{eq:helmholtz-decomposition-metric} the covariant derivative is defined with the symmetric connection \(\gamma\) whose components where defined in Eq. \eqref{eq:three-connection}. It is worth to highlight that for a torsion-free connection, the double covariant derivative of a scalar is symmetric, i.e. \(D_i D_j \hat{B} = D_j D_i \hat{B}\).

From Eq. \eqref{eq:helmholtz-decomposition-metric}, one reads that the perturbation of the metric tensor possesses four scalar components, two transverse three-dimensional (co)vectors, an a symmetric, traceless three-dimensional \(\binom{0}{2}\)-tensor.

A similar treatment can be achieved with the perturbation of the affine connection. First, we propose that the generic connection is given by the sum of an isotropic and homogeneous background connection (\(\bar{\Gamma}_{\mu}{}^{\lambda}{}_{\nu}\)) plus a \emph{small} perturbation (\(C_{\mu}{}^{\lambda}{}_{\nu}\)), i.e.
\begin{equation}
  \Gamma_{\mu}{}^{\lambda}{}_{\nu} = \bar{\Gamma}_{\mu}{}^{\lambda}{}_{\nu} + \epsilon \, C_{\mu}{}^{\lambda}{}_{\nu}.
  \label{eq:connection-perturbation}
\end{equation}
Since the perturbation \(C\) is the difference between two connections, then it is a tensor field, \(C_{\mu}{}^{\lambda}{}_{\nu} \in C^{\infty}(TM \otimes^2 T^{ * }M)\).

The scalar-vector-tensor decomposition of the \(C\)-field is given by the components, \(C_{t}{}^{t}{}_{t}\), \(C_{t}{}^{i}{}_{t}\), \(C_{i}{}^{t}{}_{t}\), \(C_{t}{}^{t}{}_{i}\), \(C_{i}{}^{t}{}_{j}\), \(C_{t}{}^{i}{}_{j}\), \(C_{j}{}^{i}{}_{t}\), and \(C_{i}{}^{j}{}_{k}\), originating a scalar component, three vector fields, three \(2\)-tensors fields and one \(3\)-tensor field. The number of contributions to the components of the symmetric part of the affine connection or to the torsion (skew-symmetric part) is shown in Tab. \ref{tab:number-components}.

In order to simplify the typesetting along the article, we shall introduce the following notation:
\begin{equation}
  \begin{aligned}
    \Sigma_{\mu\nu\lambda} & = \frac{1}{2} \left( C_{\mu\nu\lambda} + C_{\lambda\nu\mu} \right), \\
    \Lambda_{\mu\nu\lambda} & = \frac{1}{2} \left( C_{\mu\nu\lambda} - C_{\lambda\nu\mu} \right).
  \end{aligned}
  \label{eq:perturbation-tensor-symmetry}
\end{equation}
However, we have to remark that without lost of generality we might relate the original terms of the perturbation with those with lower indices as follows,
\begin{equation}
  C_{\mu t \nu} \equiv C_{\mu}{}^{t}{}_{\nu}, \quad C_{\mu i \nu} \equiv s_{i j} \,C_{\mu}{}^{j}{}_{\nu},
  \label{eq:relation-c-indices}
\end{equation}
given that the irreducible components shall be parameterised by unrelated terms in the former case, while would be related by the three-dimensional spatial metric in the later (as shown in Sec. \ref{sec:helmholtz-decomposition}).

\begin{table}[htbp]
\caption{\label{tab:number-components}Number of contributions of each term in the scalar-vector-tensor decomposition, to the symmetric and skew-symmetric components of the affine perturbation.}
\centering
\begin{tabular}{l|c|c}
Terms & Symm. (\(\Sigma\)) & Skew-symm. (\(\Lambda\))\\[0pt]
\hline
\(C_{ttt}\) & \(1\) & \(0\)\\[0pt]
\(C_{tit}\) & \(3\) & \(0\)\\[0pt]
\(C_{itt}, \, C_{tti}\) & \(3\) & \(3\)\\[0pt]
\(C_{itj}\) & \(6\) & \(3\)\\[0pt]
\(C_{ijt}, \, C_{tji}\) & \(9\) & \(9\)\\[0pt]
\(C_{ijk}\) & \(18\) & \(9\)\\[0pt]
\hline
Total components: & \(40\) & \(24\)\\[0pt]
\end{tabular}
\end{table}

A detailed analysis of the decomposition of third-order tensors is presented in Ref. \cite{landsberg11_tensor,itin21_decom_third_order_const_tensor}. However, we shall briefly discuss some of the results before proceeding to the Helmholtz decomposition. We remind the reader that in the following the discussion focus on the three-dimensional submanifold. Firstly, it should be highlighted that from the scalar-vector-tensor decomposition above, the sole novel term (not in the standard metric perturbation theory) is the corresponding to the rank \(3\) tensor, so we shall mainly focus in this term. Additionally, since these geometrical objects lie on the three-dimensional submanifold, one can use the \(s\) metric to lower the contra-variant (three-dimensional) index, reducing the problem to analysing the decomposition of completely covariant tensor field on the submanifold.

\subsection{\(SO(3,\mathbb{R})\) decomposition of the perturbation}
\label{sec:so3-decomposition}
\subsubsection{\(SO(3,\mathbb{R})\) decomposition of \(C_{ijk}\)}
\label{sec:so3-decomposition-cijk}
A general covariant \(3\)-tensor of \(GL(3,\mathbb{R})\) might be expressed as a sum of the Young irreducible components,
\begin{equation}
  \begin{aligned}
    \ydiagram{1} \otimes \ydiagram{1} \otimes \ydiagram{1}
    & =
    \left( \ydiagram{2} \oplus \ydiagram{1,1} \right) \otimes \ydiagram{1}
    \\
    & =
    \underbrace{\ydiagram{3} \oplus \ydiagram{2,1}}_{\stackrel{18}{\mathrm{symmetric}}} \oplus \underbrace{\ydiagram{2,1} \oplus \ydiagram{1,1,1}}_{\stackrel{9}{\mathrm{skew-symmetric}}}
    \\
    & =
    10_{GL_{3}} \oplus 8_{GL_{3}} \oplus 8_{GL_{3}} \oplus 1_{GL_{3}}.
  \end{aligned}
  \label{eq:young-decomposition-3-tensor}
\end{equation}
Note that if one restricts oneself to perturbations of the connection that preserve the torsion-free condition of the connection, only the first two Young diagrams are allowed, and the number of perturbative \emph{degrees of freedom} is reduced to eighteen.

The cosmological principle induces the existence of the metric \(s_{ij}\) in the three-dimensional submanifold, endowing it with an \(SO(3,\mathbb{R})\) structure. Therefore, the components in Eq. \eqref{eq:young-decomposition-3-tensor} decompose further
\begin{equation}
  \begin{aligned}
    10_{GL_{3}} & \to 7_{SO_{3}} \oplus 3_{SO_{3}}, \\
    8_{GL_{3}} & \to 5_{SO_{3}} \oplus 3_{SO_{3}}, \\
    1_{GL_{3}} & \to 1_{SO_{3}},
  \end{aligned}
  \label{eq:decomposition-gl3-to-so3}
\end{equation}
yielding the expected content: a spin \(3\) field, two spin \(2\) fields, three spin \(1\) fields and a scalar field. The above can be verified using the strategy explained in Appendix \ref{app:irreps-so3}.

\subsubsection{\(SO(3,\mathbb{R})\) decomposition of the other components}
\label{sec:so3-decomposition-other}
The components of the perturbation tensor \(C\) in the first three rows of Tab. \ref{tab:number-components} have a trivial \(SO(3,\mathbb{R})\) decomposition,
  \begin{align}
    1 & \to 1_{GL_3} \to 1_{SO_3}, \\
    3 & \to 3_{GL_3} \to 3_{SO_3}, \\
    3 & \to 3_{GL_3} \to 3_{SO_3}.
  \label{eq:so3-decomposition-first-rows}
  \end{align}

The nontrivial ones are
  \begin{align}
    6 & \to 6_{GL_3} \to 5_{SO_3} \oplus 1_{SO_3}, \\
    9 & \to 6_{GL_3} \oplus 3_{GL_3} \to 5_{SO_3} \oplus 1_{SO_3} \oplus 3_{SO_3}.
  \label{eq:so3-decomposition-nontrivial}
  \end{align}

\subsubsection{Summary: \(SO(3,\mathbb{R})\) decomposition}
\label{sec:so3-decomposition-summary}
The results from the last sections is summarised in Tab. \ref{tab:so3-decomposition-summary}.

\begin{widetext}

\begin{table}[htbp]
\caption{\label{tab:so3-decomposition-summary}Summary of irreducible representations of \(SO(3,\mathbb{R})\) obtained from the irreducible components of the affine perturbation tensor \(C\), in terms of its symmetric (\(\Sigma\)) and skew-symmetric (\(\Lambda\)) parts.}
\centering
\begin{tabular}{c|c|c|c}
Term & Components & \(GL(3,\mathbb{R})\) & \(SO(3,\mathbb{R})\)\\[0pt]
\hline
\(\Sigma_{ttt}\) & \(1_s\) & \(1_{GL_3}\) & \(1_{SO_3}\)\\[0pt]
\(\Sigma_{tit}\) & \(3_s\) & \(3_{GL_3}\) & \(3_{SO_3}\)\\[0pt]
\(\Sigma_{tti}\) & \(3_s\) & \(3_{GL_3}\) & \(3_{SO_3}\)\\[0pt]
\(\Lambda_{tti}\) & \(3_a\) & \(3_{GL_3}\) & \(3_{SO_3}\)\\[0pt]
\(\Sigma_{itj}\) & \(6_s\) & \(6_{GL_3}\) & \(5_{SO_3} \oplus 1_{SO_3}\)\\[0pt]
\(\Lambda_{itj}\) & \(3_a\) & \(3_{GL_3}\) & \(3_{SO_3}\)\\[0pt]
\(\Sigma_{tij}\) & \(9_s\) & \(6_{GL_3} \oplus 3_{GL_3}\) & \(5_{SO_3} \oplus 1_{SO_3} \oplus 3_{SO_3}\)\\[0pt]
\(\Lambda_{tij}\) & \(9_a\) & \(6_{GL_3} \oplus 3_{GL_3}\) & \(5_{SO_3} \oplus 1_{SO_3} \oplus 3_{SO_3}\)\\[0pt]
\(\Sigma_{ijk}\) & \(18_s\) & \(10_{GL_3} \oplus 8_{GL_3}\) & \(7_{SO_3} \oplus 3_{SO_3} \oplus 5_{SO_3} \oplus 3_{SO_3}\)\\[0pt]
\(\Lambda_{ijk}\) & \(9_a\) & \(6_{GL_3} \oplus 3_{GL_3}\) & \(5_{SO_3} \oplus 1_{SO_3} \oplus 3_{SO_3}\)\\[0pt]
\end{tabular}
\end{table}
\end{widetext}

\subsection{Helmholtz decomposition of the perturbation}
\label{sec:helmholtz-decomposition}
The final piece in the standard cosmological perturbation theory is the Helmholtz decomposition of the perturbative components from which originates the results in Eq. \eqref{eq:helmholtz-decomposition-metric}. In this section, we develop the Helmholtz decomposition for the perturbation tensor \(C\).

Before presenting explicit formulas for the Helmholtz decomposition of the field \(C\), we would argue that the Helmholtz decomposition is equivalent to the decomposition of representations of \(SO(3,\mathbb{R})\) into irreducible representations of \(SO(2,\mathbb{R})\) (some details of this decomposition are presented in Appendix \ref{app:so3-to-so2}).

Consider the components of the metric perturbation, \(h_{\mu\nu}\), and their Helmholtz decomposition as shown in Eq. \eqref{eq:helmholtz-decomposition-metric}.

The component \(h_{tt}\) is a scalar of \(SO(3,\mathbb{R})\), i.e. \(h_{tt} = 1_{SO_3}\), but the trivial representation of \(SO(3,\mathbb{R})\) yields the trivial representation of \(SO(2,\mathbb{R})\),
\begin{equation}
  h_{tt} = 1_{SO_3} \to 1_{SO_2} = \hat{E}.
  \label{eq:helmholtz-htt}
\end{equation}
Similarly, the component \(h_{ti}\) is a vector of \(SO(3,\mathbb{R})\), i.e. \(h_{ti} = 3_{SO_3}\). However, the representation \(3_{SO_3}\) decomposes non-trivially into irreducible representations of \(SO(2,\mathbb{R})\) according to Eq. \eqref{eq:so3-to-so2}, yielding
\begin{equation}
  h_{ti} = 3_{SO_3} \to 1_{SO_2} \oplus 2_{SO_2} = D_i \hat{F} + \hat{G}_i.
  \label{eq:helmholtz-hti}
\end{equation}
Finally, the component \(h_{ij}\) is a symmetric \(2\)-tensor of \(SO(3,\mathbb{R})\), i.e. \(h_{ij} = 5_{SO_3} + 1_{SO_3}\). Then, its decomposition into irreducible representations of \(SO(2,\mathbb{R})\) yields
\begin{equation}
  \begin{aligned}
    h_{ij} & = 5_{SO_3} + 1_{SO_3} \\
           & \to 1_{SO_2} \oplus 2_{SO_2} \oplus 2_{SO_2} \oplus 1_{SO_2} \\
           & = \left( D_i D_j - \frac{s_{ij}}{3} D^{2} \right) \hat{B} + 2 D_{(i} \hat{C}_{j)} + \hat{D}_{ij} + \frac{s_{ij}}{3} \hat{A}.
  \end{aligned}
  \label{eq:helmholtz-hij}
\end{equation}
In the last line of Eq. \eqref{eq:helmholtz-hij}, \(D^2\) should be understood as \(s^{ij} D_i D_j\).

From the information in Tab. \ref{tab:so3-decomposition-summary} and Appendix \ref{app:so3-to-so2}, one can read the number of fields in the Helmholtz decomposition of the affine connection, which are summarized in Tab. \ref{tab:count-helmholtz-fields}.

\begin{table}[htbp]
\caption{\label{tab:count-helmholtz-fields}Number of scalars (\(T_0\)), vectors (\(T_1\)), \(2\)-tensors (\(T_2\)) and \(3\)-tensors (\(T_3\)), obtained from the Helmholtz decomposition of the irreducible components of the affine connection.}
\centering
\begin{tabular}{c|c|c|c|c}
Component & \(T_0\) & \(T_1\) & \(T_2\) & \(T_3\)\\[0pt]
\hline
\(1_s\) & \(1\) &  &  & \\[0pt]
\(3_s\) & \(1\) & \(1\) &  & \\[0pt]
\(3_a\) & \(1\) & \(1\) &  & \\[0pt]
\(6_s\) & \(2\) & \(1\) & \(1\) & \\[0pt]
\(9_a\) & \(3\) & \(2\) & \(1\) & \\[0pt]
\(18_s\) & \(4\) & \(4\) & \(2\) & \(1\)\\[0pt]
\end{tabular}
\end{table}

Therefore, we can now write the Helmholtz decomposition of the tensor \(C\), following the spirit of the Eqs. \eqref{eq:helmholtz-htt}, \eqref{eq:helmholtz-hti} and \eqref{eq:helmholtz-hij}. 

The component \(\Sigma_{ttt}\) is an \(SO(3,\mathbb{R})\)-scalar, and therefore cannot be decomposed further. Hence,
\begin{equation}
  \Sigma_{ttt} = 1_{SO_3} \to 1_{SO_2} = A.
  \label{eq:helmholtz-sttt}
\end{equation}

The components \(\Sigma_{tit}\), \(\Sigma_{tti}\), \(\Lambda_{tti}\) and \(\Lambda_{itj}\) are \(SO(3,\mathbb{R})\)-vectors, and each of them decomposes into a scalar (longitudinal) component and a transverse vector,
\begin{equation}
  3_{SO_3} \to 2_{SO_2} \oplus 1_{SO_2},
  \label{eq:helmholtz-so3-vector}
\end{equation}
written explicitly as,
\begin{align}
  \label{eq:helmholtz-stit}
  \Sigma_{tit} & = D_i B + C_i,\\
  \label{eq:helmholtz-stti}
  \Sigma_{tti} & = D_i D + E_i, \\
  \label{eq:helmholtz-ltit}
  \Lambda_{tti} & = D_i \widetilde{B} + \widetilde{C}_i,\\
  \label{eq:helmholtz-ltti}
  \Lambda_{itj} & = \sqrt{s} \, \varepsilon_{ijk} \, s^{kl} (D_l \widetilde{D} + \widetilde{E}_l) ,
\end{align}
where \(\mathcal{X}^{i}\), defined as \(\mathcal{X}^i = s^{ij} \mathcal{X}_j\) for \(\mathcal{X}_i \in \left\{ C_i, E_i, \widetilde{C}_i, \widetilde{E}_i \right\}\), satisfies \(D_i \mathcal{X}^i = 0\). Hereon, whenever it is said that a tensor is \emph{transverse} it should be understood as the later relation. In order to simplify notations, perturbations originating from the antisymmetric component of the connection $(\Lambda_{ijk})$ will be denoted with a tilde, whereas perturbations derived from the symmetric component of the connection $(\Sigma_{ijk})$ will be expressed without a tilde.

The component \(\Sigma_{itj}\) decomposes just like the perturbation \(h_{ij}\) of the metric tensor, because it contains both the symmetric and traceless rank \(2\) tensor (\(5_{SO_3}\)) and its trace (\(1_{SO_3}\)). The explicit Helmholtz decomposition is given by
\begin{equation}
  \Sigma_{itj} = \frac{s_{ij}}{3} F + \left( D_i D_j - \frac{s_{ij}}{3} D^2 \right) G + 2 D_{(i} H_{j)} + I_{ij},
  \label{eq:helmholtz-sitj}
\end{equation}
where \(I_{ij}\) is symmetric and traceless, and both \(H_i\) and \(I_{ij}\) are transverse.

The components \(\Sigma_{tij}\) and \(\Lambda_{tij}\), which correspond to the \(9_s\) and \(9_a\) representations of \(GL(3,\mathbb{R})\) respectively, are obtained as the sum of a \(5_{SO_3}\), \(3_{SO_3}\) and \(1_{SO_3}\), whose Helmholtz decomposition is given explicitly by
\begin{equation}
  \begin{aligned}
    \Sigma_{tij} & = \sqrt{s} \varepsilon_{ijk} s^{kl} \left( D_l J + K_l \right) + \frac{s_{ij}}{3} L \\
    & + \left( D_i D_j - \frac{s_{ij}}{3} D^2 \right) M + 2 D_{(i} N_{j)} + O_{ij},
  \end{aligned}
  \label{eq:helmholtz-stij}
\end{equation}
and
\begin{equation}
  \begin{aligned}
    \Lambda_{tij} & = \sqrt{s} \varepsilon_{ijk} s^{kl} \left( D_l \tilde{J} + \tilde{K}_l \right) + \frac{s_{ij}}{3} \tilde{L} \\
    & + \left( D_i D_j - \frac{s_{ij}}{3} D^2 \right) \tilde{M} + 2 D_{(i} \tilde{N}_{j)} + \tilde{O}_{ij},
  \end{aligned}
  \label{eq:helmholtz-ltij}
\end{equation}
where all the tensor quantities are symmetric, traceless and transverse.

The components \(\Sigma_{ijk}\) and \(\Lambda_{ijk}\), corresponding to the \(18_s\) and \(9_{a}\) representation of \(GL(3,\mathbb{R})\) decompose as follows,
\begin{widetext}
\begin{equation}
  \begin{aligned}
    \Sigma_{ijk}
    & =
    \frac{3}{5} \left( s_{(ij} D_{k)} P + s_{(ij} Q_{k)} \right) +
    \left( D_{(i} D_j D_{k)}
    - \frac{2}{5} D^2  s_{(ij} D_{k)}
    - \frac{1}{5} s_{(ij} D_{k)} D^2 \right) R
    \\
    & \quad
    + D_{(i} D_j S_{k)}
    - \frac{1}{5} D^2 s_{(ij} S_{k)}
    - \frac{1}{5} s_{(ij} D^m D_{k)} S_m
    + D_{(i} T_{jk)} + U_{ijk}
    + \frac{1}{2} \sqrt{s} s^{pq} \left( \varepsilon_{ijp} \delta_k^r + \varepsilon_{kjp} \delta_i^r \right)
    \\
    & \qquad \left[ \left(D_q D_r - \frac{1}{3} s_{qr} D^2 \right) V + 2 D_{(q} W_{r)} + X_{qr} + \sqrt{s} \varepsilon_{qrm} s^{mn} (D_n Y + Z_n) \right],
  \end{aligned}
  \label{eq:helmholtz-sijk}
\end{equation}
and
\begin{equation}
  \begin{aligned}
    \Lambda_{ijk}
    & = \sqrt{s} \varepsilon_{ijk} \tilde{A} + \frac{1}{2} \sqrt{s} s^{pq} \left( 2 \varepsilon_{ikp} \delta_j^r + \varepsilon_{ijp} \delta_k^r - \varepsilon_{kjp} \delta_i^r \right) \\
    & \quad \left[ \left( D_q D_r - \frac{1}{3} s_{qr} D^2 \right) \tilde{V} + 2 D_{(q} \tilde{W}_{r)} + \tilde{X}_{qr} + \sqrt{s} \varepsilon_{qrm} s^{mn} (D_n \tilde{Y} + \tilde{Z}_n) \right], 
  \end{aligned}
  \label{eq:helmholtz-lijk}
\end{equation}
\end{widetext}
where all the tensor quantities, as in the previous cases, are symmetric, traceless and transverse. 

In summary, the perturbations have been decomposed into sixteen scalars, four pseudoscalars, ten transverse vectors, five transverse pseudovectors, six transverse and traceless (TT) 2-tensors and finally one TT 3-tensor. This decomposition becomes crucial when examining the dynamics of a parity-preserving theory. For instance, when expanding the action to the second order of perturbations, the absence of terms like $AJ$ is certain because it would violate parity, ensuring that scalars and pseudoscalars do not mix. Conversely, terms like $\epsilon_{ijk}(\partial_i C_j)K_k$ can appear and therefore leading to the mixing of vector and pseudovector perturbations, highlighting the potential for interaction in this context.

\begin{table}[htbp]
\centering
\begin{tabular}{l|l}
Scalars & $A,B,D,F,G,L,M,P,R,V,\tilde A,\tilde B,\tilde L,\tilde M,\tilde V,\tilde Y$\\[0.05cm]
Pseudoscalars & $J,Y,\tilde D,\tilde J$\\[0.05cm]
Vectors & $C_i,E_i,H_i,N_i,Q_i,S_i,W_i,\tilde C_i,\tilde N_i,\tilde W_i$\\[0.05cm]
Pseudovectors & $K_i,Z_i,\tilde E_i,\tilde K_i,\tilde Z_i$\\[0.05cm]
2-tensor & $I_{ij},O_{ij},T_{ij},X_{ij},\tilde O_{ij}, \tilde X_{ij}$ \\[0.05cm]
3-tensor & $U_{ijk}$\\[0.05cm]
\end{tabular}
\end{table}

\section{Gauge invariance}
\label{sec:gauge}
The next step in the analysis is to find the set of adequate variables to describe the perturbations, i.e. combinations of the perturbations that are invariant under general coordinate transformations \cite{weinberg08_cosmol,poisson14_gravit,dodelson20_moder}.

Following a similar procedure than for the metric perturbations, it is straightforward to show that the \emph{gauge} transformation of the affine perturbations is given by
\begin{equation}
  \begin{aligned}
    \delta C_{\mu}{}^{\lambda}{}_{\nu}
    & = \pounds_{\xi} \Gamma_{\mu}{}^{\lambda}{}_{\nu} \\
    & = \xi^\sigma R_{\sigma \mu}{}^{\lambda}{}_{\nu} + \nabla_\mu \nabla_\nu \xi^\lambda 
      - \nabla_\mu ( T_\nu{}^\lambda{}_\sigma \xi^\sigma ).
  \end{aligned}
  \label{eq:transformation-perturbation}
\end{equation}

After a simple but rather lengthy calculation, one obtains the transformation rules for the components of the perturbation,
\begin{equation}
  \delta C_{t}{}^{t}{}_{t} = \ddot{\xi}^{t},
  \label{eq:dCttt}
\end{equation}
\begin{equation}
  \delta C_{t}{}^{i}{}_{t} = \ddot{\xi}^i + 2 h \dot{\xi}^i,
  \label{eq:dCtit}
\end{equation}
\begin{equation}
  \delta C_{t}{}^{t}{}_{i} = D_i \dot{\xi}^{t} + g s_{ij} \dot{\xi}^j - (h-p) D_i \xi^{t},
  \label{eq:dCtti}
\end{equation}
\begin{equation}
  \delta C_{i}{}^{t}{}_{t} = 
  D_{i}\dot{\xi}^{t} - (h+p) D_{i}{\xi^{t}} +  g s_{i j} \dot{\xi}^j,
  \label{eq:dCitt}
\end{equation}
\begin{equation}
  \begin{aligned}
    \delta C_{i}{}^{t}{}_{j} & =
    D_i D_j \xi^{t} + 2 g s_{k(i} D_{j)} \xi^k \\
    & \quad - g q \left( 2 s_{kj} S_i{}^k{}_l + s_{ki} S_j{}^k{}_l + s_{kl} S_i{}^k{}_j \right) \xi^l
    \\
    & \quad + s_{i j} (\dot{g} \xi^{t} - g \dot{\xi}^{t}) - q S_i{}^k{}_j D_k \xi^{t},
  \end{aligned}
  \label{eq:dCitj}
\end{equation}
\begin{equation}
  \delta C_{t}{}^{j}{}_{i} = D_i \dot{\xi}^{t} + \delta_i^j \partial_{t}( (h-p) \xi^{t} ) - q \, S_i{}^j{}_k \dot{\xi}^k,
  \label{eq:dCtji}
\end{equation}
\begin{equation}
  \delta C_{i}{}^{j}{}_{t} = D_i \dot{\xi}^{t} + \delta_i^j \partial_{t}( (h+p) \xi^{t} ) + q \, S_i{}^j{}_k \dot{\xi}^k,
  \label{eq:dCijt}
\end{equation}
\begin{equation}
  \begin{aligned}
    \delta C_i{}^k{}_j
    & =  D_i D_j \xi^k + \kappa \left( \delta_l^k s_{ij} - \delta_i^k s_{lj} \right) \xi^l - g \, s_{ij} \dot{\xi}^k \\
    & \quad + \dot{q} S_i{}^k{}_j \xi^{t} + (h+p) \delta_i^k D_j \xi^{t} + (h-p) \delta_j^k D_i \xi^{t} \\
    & \quad - q \left( S_i{}^l{}_j D_l \xi^k + 2 S_l{}^k{}_{[i} D_{j]} \xi^l \right)
  \end{aligned}
  \label{eq:dCikj}
\end{equation}

In order to read the transformation rules of the components of the perturbations under coordinate transformations, we have to decompose the spatial generator of the transformation into its longitudinal and transverse components,
\begin{equation}
  \xi^i \to D^i \psi + \zeta^i \text{ where } D_i \zeta^i = 0,
  \label{eq:split-generator}
\end{equation}
and when possible to separate the symmetric and skew-symmetric part of the transformations, to identify the variation of the irreducible components of \(\Sigma\)'s and \(\Lambda\)'s.

From Eqs. \eqref{eq:helmholtz-sttt} and \eqref{eq:dCttt} one obtains directly that
\begin{equation}
  \delta A = \ddot{\xi}^t,
  \label{eq:dA}
\end{equation}
since \(\delta C_{ttt} = \delta A\).

Similarly, Eqs. \eqref{eq:helmholtz-stit} and \eqref{eq:dCtit} yield the variations,
\begin{align}
  \label{eq:dB}
  \delta B & = \ddot{\psi} + 2 h \dot{\psi}, \\
  \label{eq:dC}
  \delta C^i & = \ddot{\zeta}^{i} + 2 h \dot{\zeta}^{i}.
\end{align}

Next, one has to combine the transformations in Eqs. \eqref{eq:dCtti} and \eqref{eq:dCitt} to get the transformations of \(\Sigma_{t}{}^{t}{}_{i}\) and \(\Lambda_{t}{}^{t}{}_{i}\). From these combinations one reads the variations
\begin{align}
  \label{eq:dD}
  \delta D & = \dot{\xi}^t - h \xi^t + g \dot{\psi},\\
  \label{eq:dE}
  \delta E_i & = g \, s_{ij} \dot{\zeta}^j \\
  \label{eq:dBt}
  \delta \tilde{B} & = p \, \xi^t , \\
  \label{eq:dCt}
  \delta \tilde{C}^i & = 0.
\end{align}

The transformation in Eq. \eqref{eq:dCitj} has to be separated into symmetric and skew-symmetric parts. The skew-symmetric part yields directly the variations associated to the irreducible components of \(\Lambda_{itj}\), as follows
\begin{align}
  \label{eq:dDt}
  \delta \tilde{D} & = q \, \xi^{t}, \\
  \label{eq:dEt}
   \delta \tilde{E}_i & = 0.
\end{align}
However, the symmetric part has to be decomposed further to take out its trace. Hence, one gets the variations
\begin{align}
  \label{eq:dF}
  \delta F & = 3 \dot{g} \xi^t - 3 g \dot{\xi}^t + D^2 (\xi^t + 2 g \psi), \\
  \label{eq:dG}
  \delta G & = \xi^t + 2 g \psi, \\
  \label{eq:dH}
  \delta H_i & = g \zeta_i, \\
  \label{eq:dI}
  \delta I_{ij} & = 0.
\end{align}

From Eqs. \eqref{eq:dCtji}  and \eqref{eq:dCijt} one would get the variation of the irreducible components of \(\Sigma_{tji}\) and \(\Lambda_{tji}\), after considering their symmetrisation and the skew-symmetrisation. The variations obtained from \(\Sigma_{tji}\) are 
\begin{align}
  \label{eq:dJK}
  \delta \left( D_l J + K_l \right) & = - \frac{1}{2 \sqrt{s}} s_{l k} \epsilon^{k i j} D_i \zeta_j, \\
  \label{eq:dL}
  \delta L & = 3 ( \dot{h} \xi^t + h \dot{\xi}^t ) + D^2 \dot{\psi}, \\
  \label{eq:dM}
  \delta M & = \dot{\psi}, \\
  \label{eq:dN}
  \delta N_i & = \frac{1}{2} \dot{\zeta}_i , \\
  \label{eq:dO}
  \delta O_{ij} & = 0,
\end{align}
while the ones obtained from the skew-symmetric are
\begin{align}
  \label{eq:dJt}
  \delta \tilde{J} & = q \dot{\psi}, \\
  \label{eq:dKt}
  \delta \tilde{K}_{i} & = q s_{ij} \dot{\zeta}^j, \\
  \label{eq:dLt}
  \delta \tilde{L} & = - 3 ( \dot{p} \xi^t + p \dot{\xi}^t ), \\
  \label{eq:dMt}
  \delta \tilde{M} & = 0, \\
  \label{eq:dNt}
  \delta \tilde{N}_{i} & = 0, \\
  \label{eq:dOt}
  \delta \tilde{O}_{ij} & = 0.
\end{align}

Finally, from Eq. \eqref{eq:dCikj} one obtains the variations for the irreducible components of \(\Sigma_{ikj}\) and \(\Lambda_{ikj}\). Following a procedure similar to the previous one, one obtains that the variations of the components of \(\Sigma_{ikj}\) are
\begin{align}
  \label{eq:dP}
  \delta P & = D^2 \psi + \frac{4}{3} \kappa \psi + \frac{10}{3} h \xi^{t} - \frac{5}{3} g \dot{\psi}, \\
  \label{eq:dQ}
  \delta Q_i & = \frac{2}{3} \kappa \zeta_i + \frac{1}{3} D^2 \zeta_i - \frac{5}{3} g 
\dot{\zeta}_i, \\
  \label{eq:dR}
  \delta R & = \psi, \\
  \label{eq:dS}
  \delta S_i & = \zeta_i, \\
  \label{eq:dT}
  \delta T_{ij} & = 0, \\
  \label{eq:dU}
  \delta U_{ikj} & = 0 \\
  \label{eq:dV}
  \delta  V & = 0, \\
  \label{eq:dW}
  \delta W_i & = \frac{1}{3} \sqrt{s} \epsilon_{ijk} D^j \zeta^k, \\
  \label{eq:dX}
  \delta X_{ij} & = 0, \\
  \label{eq:dY}
  \delta Y & = \frac{4}{3} \kappa \psi - \frac{2}{3} h \xi^{t} - \frac{2}{3} g \dot{\psi}, \\
  \label{eq:dZ}
  \delta Z_i & = \frac{2}{3} \kappa \zeta_i + \frac{1}{3} D^2 \zeta_i - \frac{2}{3} g \dot{\zeta}_i.
\end{align}
While the components of the \(\Lambda_{ikj}\) vary as follows,
\begin{align}
  \label{eq:dAt}
  \delta \tilde{A} & = \dot{q} \xi^t + \frac{1}{3} q D^2 \psi, \\
  \label{eq:dVt}
  \delta \tilde{V} & = \frac{4}{3} q \psi, \\
  \label{eq:dWt}
  \delta \tilde{W}_i & = \frac{2}{3} q \zeta_i, \\
  \label{eq:dXt}
  \delta \tilde{X}_{ij} & = 0, \\
  \label{eq:dYt}
  \delta \tilde{Y} & = \frac{2}{3} p \xi^t, \\
  \label{eq:dZt}
  \delta \tilde{Z}_i & = 0.
\end{align}

In addition to the obvious gauge-invariant tensors of rank two and three,
we observe that the perturbation $V$ for the symmetric part and the perturbations $\tilde M,\tilde C_i,\tilde E_i,\tilde B_i,\tilde Z_i$ of the antisymmetric part are also gauge invariant. This observation implies that $24$ components of the perturbations are gauge invariant, leaving $40$ components whose transformations are detailed in the preceding equations. Specifically, $29$ components pertain to the symmetric part of the perturbation, while 11 components belong to the antisymmetric part. Of course the gauge freedom, can be used to reduce them to $36$ components. In the upcoming discussion, we will elaborate on different potential gauges by setting specific variables to zero. In our context, this flexibility will be exclusively applied to the gravitational sector, as we are not taking into account matter fields.

We can leverage the freedom associated with the time coordinate to eliminate one of the seven perturbations $(G,P,Y,\tilde A,\tilde B, \tilde D,\tilde Y)$. Additionally, the transformation in the space coordinate allows us to choose $\psi$ strategically, eliminating one of these five perturbations $(F,G,R,\tilde A,\tilde V)$. Finally, the freedom associated with $\zeta^i$ enables the removal of one perturbation from the set $(H_i,S_i,W_i,\tilde W_i, D_iJ+K_i)$ giving a total of $165$ different gauges.\footnote{While we have considered a generic spacetime, it is important to note that in specific cases, the number of invariant quantities may vary. For instance, if the solution is torsion-free, i.e., $p=q=0$, this condition implies the existence of new invariant quantities.} We see that one particularly interesting gauge will be to set $J=K_i=0$ with $\zeta^i$, eq. (\ref{eq:dJK}), because $J$ is a pseudoscalar which transforms via a vector, producing therefore a mixing term. 

Finally, let us notice that for a background with zero torsion, indicating $p=q=0$, the following fields remain invariant $(\tilde{A},\tilde{B},\tilde{D},\tilde{J},\tilde{L},\tilde{V},\tilde{Y},\tilde{K}_i,\tilde{W}_i)$. This implies the invariance of the tensor $\Lambda_{ijk}$. Such invariance is a consequence of the well-known Stewart-Walker lemma, which asserts that a tensor is gauge-invariant if it vanishes in the background \cite{Stewart:1974uz}.

\section{Discussion and Conclusions}
\label{sec:org6977b26}

In this paper, we have assumed that the structure of our spacetime is solely characterized by the connection. Within this framework, a pivotal undertaking is the development of a comprehensive cosmological perturbation formalism. By presuming a background consistent with a homogeneous and isotropic Universe, the perturbations manifest through the connection. Given the absence of a metric, we contend with a total of $64$ fields associated with the connection. Despite the non-tensorial nature of the connection, the perturbation itself is a tensor, and we have effectively decomposed it into two distinct parts: a symmetrical portion and an antisymmetrical component, corresponding to torsion. This decomposition facilitates the further breakdown of the perturbation into $40$ components for the symmetrical part and $24$ components for the antisymmetric part. It is noteworthy that, in line with General Relativity, where torsion is assumed to be zero, we retrieve the standard $40$ components.

The various fields associated with the connection undergo decomposition into irreducible components, akin to the Helmholtz decomposition in the conventional cosmological framework. Furthermore, mirroring the principles of general relativity, we use the freedom to perform a coordinate transformation to simplify our problem. We have meticulously derived the complete set of transformations, yielding a total of $165$ potential gauges in the general case. In a subsequent paper, we will delve into specific problems and explore which gauges prove advantageous for simplifying algebraic calculations.

Finally, we observe that when the antisymmetric part of the connection is zero in the background, its perturbation becomes gauge invariant—a manifestation of the Stewart-Walker lemma.

In a subsequent paper, we will investigate the dynamics of a specific model within this formalism.

\begin{acknowledgments}
OCF acknowledge the financial support received by ANID PIA/APOYO~AFB230003  (Chile) and FONDECYT Regular No.~1230110 (Chile). The work of MML has also been supported by the ANID ``Doctorado Nacional'' grant No.~2023-21231762 (Chile), by USM ``Programa de Iniciación a la Investigación Científica'', and by ESPOL ``Programa de Ayuda Económica para fomentar la producción científica de profesores no titulares que se encuentren haciendo estudios de postgrado en el extranjero''. The work of R.G. is supported by ANID FONDECYT Regular No.~1220965 (Chile). The work of M. Rozas-Rojas is supported by CONVENIO DE BECA DE DOCTORADO No.~037/2023.
\end{acknowledgments}

\appendix

\section{Irreducible representations of \(SO(3,\mathbb{R})\) using Young tableaux}
\label{app:irreps-so3}
In theoretical physics the Young tableaux techniques to calculate the irreducible representations of groups \(SU(N,\mathbb{C})\) are extensively used \cite{Georgi:1982jb}. However, the computation of irreducible representations of \(SO(N,\mathbb{R})\) does require additional strategies, due to the possibility of extracting traces with respect to its invariant symmetric \(2\)-tensor.

For the particular case of the group \(SO(3,\mathbb{R})\)---whose algebra is \(\mathfrak{so}(3) \simeq \mathfrak{su}(2)\)---, one can use the local isomorphism with its covering group \(SU(2,\mathbb{C})\) to find its irreducible representations through the Young tableaux technique.

The \emph{trick} is to build the representations starting from the products of the three-dimensional representation of \(SU(2,\mathbb{C})\), which is depicted by the tableau \(\ydiagram{2}\).

Hence, the tensor product of two fundamental representations of \(SO(3,\mathbb{R})\) is obtained by
\begin{equation}
  \begin{aligned}
    \ydiagram[*(gray!80)]{2}
    \otimes
    \ydiagram[*(gray!20)]{2}
    & =
    \ydiagram[*(gray!20)]{2+2} *[*(gray!80)]{4}
    \oplus
    \ydiagram[*(gray!20)]{2+1,1} *[*(gray!80)]{3,1}
    \oplus
    \ydiagram[*(gray!20)]{2+0,2} *[*(gray!80)]{2,2}
    \\
    & =
    \ydiagram{4} \oplus \ydiagram{2} \oplus \bullet
    \\
    & =
    5 \oplus 3 \oplus 1,
  \end{aligned}
  \label{eq:3-times-3-so3}
\end{equation}
which correspond to the trace-less symmetric, skew-symmetric and trace components respectively.

Similarly, a generic rank three tensor of \(SO(3,\mathbb{R})\) decomposes into the following irreducible parts,
\begin{equation}
  \begin{aligned}
    & \Big(
    \ydiagram[*(gray!80)]{2}
    \otimes
    \ydiagram[*(gray!20)]{2}
    \Big)
    \otimes
    \ydiagram{2}
    \\
    & \quad =
    \Big(
      \ydiagram{4} \oplus \ydiagram{2} \oplus \bullet
    \Big)
    \otimes
    \ydiagram{2}
    \\
    & \quad =
    \ydiagram{6} \oplus \ydiagram{4} \oplus \ydiagram{2}
    \\
    & \qquad
    \oplus \ydiagram{4} \oplus \ydiagram{2} \oplus \bullet
    \\
    & \qquad
    \oplus \ydiagram{2}
    \\
    & \quad =
    \left( 7 \oplus 5 \oplus 3 \right) \oplus \left( 5 \oplus 3 \oplus 1 \right) \oplus 3,
  \end{aligned}
  \label{eq:3-times-3-times-3-so3}
\end{equation}

\section{From \(SO(3,\mathbb{R})\) to \(SO(2,\mathbb{R})\)}
\label{app:so3-to-so2}
The irreducible representations of \(SO(3,\mathbb{R})\) (over real vector spaces), have dimension \((2 \ell + 1)\) for \(\ell \in \mathbb{N}\), and the (conjugacy) classes of the group are constituted by all elements that \emph{rotate} by a certain angle \(\theta\). The character of the (conjugacy) class of elements characterised by the angle \(\theta\), in the irreducible representation labelled by \(\ell\) is

\begin{equation}
  \chi^{(\ell)}_{SO_3}(\theta) = 1 + 2 \left( \cos(\theta) + \cos(2 \theta) + \cdots + \cos(\ell \theta) \right).
  \label{eq:character-so3}
\end{equation}

The all irreducible representations of \(SO(2,\mathbb{R})\) (over real vector spaces), with the exception of the trivial representation, are two-dimensional and labelled by \(\ell \in \mathbb{N}^{ * }\). Explicitly, an element characterised by the parameter \(\theta\) in the irreducible representation labelled by \(\ell\) has the form
\begin{equation}
  D^{(\ell)}(g(\theta)) =
  \begin{pmatrix}
    \cos(\ell \theta) & - \sin(\ell \theta) \\
    \sin(\ell \theta) &   \cos(\ell \theta)
  \end{pmatrix},
  \label{eq:so2-representation-l}
\end{equation}
and its character is
\begin{equation}
  \chi^{(\ell)}_{SO_2}(\theta) = 2 \cos(\ell \theta).
  \label{eq:character-so2}
\end{equation}

The embedding of \(SO(2,\mathbb{R})\) into \(SO(3,\mathbb{R})\) yields the decomposition of irreducible representations of the latter in terms of irreducible representations of the former as follows,
\begin{equation}
  (2\ell + 1)_{SO_3} = 1_{SO_2} + \underbrace{2_{SO_2} + \cdots + 2_{SO_2}}_{\ell \text{ times}},
  \label{eq:so3-to-so2}
\end{equation}
where there is a representations \(2_{SO_2}\) for each value \(\ell^{\prime} \in [1, 2, \cdots, \ell]\).

% \bibliography{References}

\end{document}